\documentclass[12pt]{article}
\def\ypiz_2gg{$\pi^0 \rightarrow \gamma\gamma$}
\newcommand {\charex} {$K^+\mathrm{Xe} \rightarrow K^0 p \mathrm{Xe}'$}

\begin {document}
\title
{Observation of a baryon resonance with positive strangeness in
$K^+$ collisions with Xe nuclei\footnote{Based on a talk at Session of Nuclear
Division of Russian Academy of Sciences, Dec. 3, 2002. 
Published in Yad. Fiz. 66, 1763-1766 (2003); Phys. At. Nucl. 66, 1715--1718 (2003).
}}
\author{
DIANA Collaboration\\
V.V. Barmin$^a$, V.S. Borisov$^a$, G.V. Davidenko$^a$, 
A.G. Dolgolenko$^{a,}$\thanks{Corresponding author. E-mail address:
dolgolenko@vitep1.itep.ru.},\\
C. Guaraldo$^b$, I.F. Larin$^a$, V.A. Matveev$^a$, C. Petrascu$^b$,\\
V.A. Shebanov$^a$, N.N. Shishov$^a$, L.I. Sokolov$^a$,\\
and G.K. Tumanov$^a$\\
\normalsize {$^a$ \it Institute of Theoretical and Experimental Physics,
Moscow 117259, Russia}\\
\normalsize {$^b$ \it Laboratori Nazionali di Frascati dell' INFN,
C.P. 13-I-00044 Frascati, Italy}\\
}                                          % authors
\date {    }
\maketitle

\begin{abstract}
The status of our investigation of low-energy $K^+$Xe collisions in 
the Xenon bubble chamber DIANA is reported. In the charge-exchange 
reaction \charex, the spectrum of $K^0 p$ effective mass shows a
resonant enhancement with $M = 1539 \pm 2$ MeV/c$^2$ and $\Gamma \le 9$
MeV/c$^2$. The statistical significance of the enhancement is near
$4.4\sigma$. The mass and width of the observed resonance are consistent
with expectations for the lightest member of the anti-decuplet of 
exotic pentaquark baryons, as predicted in the framework of the chiral 
soliton model.
\end{abstract}
%   {\it PACS:} 14.60.Pq; 14.60.Fg \\
%   {\it Keywords:} Neutrino oscillations; $\nu_e$ appearance\\

\newpage
     This paper reports an investigation of low-energy $K^+$--nucleus
collisions aimed at testing the hypothesis of an anti-decuplet of exotic
pentaquark baryons, as proposed by Diakonov, Petrov and Polyakov \cite{diakonov} 
in the framework of the chiral soliton model. Assuming 
that the known nucleon resonance $P_{11}(1710)$ belongs to the 
hypothesized anti-decuplet, predictions for the masses and decay widths 
of its other members have been obtained \cite{diakonov}. Of these,
the lightest is the exotic pentaquark state $uudd\bar{s}$ with mass near 
1530 MeV/c$^2$ and decay width $\Gamma < 15$ MeV/c$^2$. This spin-1/2 
isospin-zero state with positive strangeness, referred to as the $Z^+$ 
baryon, is expected to decay to $K^0 p$ and $K^+ n$.

     Two different methods may be employed when searching for formation 
of the $Z^+$ baryon in $K^+$Xe collisions. The first and straightforward
approach is to analyze the effective mass of the $K^0 p$ system in the 
reaction \charex\ for the $Z^+$ peak near 1530 MeV/c$^2$. The second
approach is to measure the cross sections for formation of the final 
states $K^0 p$ and $K^+ n$ as functions of $K^+$ momentum. Given a 
target nucleon bound in the Xe nucleus, formation of the $Z^+$ baryon 
should manifest itself as an enhancement of partial 
cross sections of elementary processes  $K^+n \rightarrow K^+n$  and 
$K^+n \rightarrow K^0 p$  at $K^+$ momentum near 480 MeV/c. Measurements
of these partial cross sections in the interval $400 < P_{K^+} < 700$ 
MeV/c are in progress, and should finally reach a statistical accuracy 
of (3--6)\% with $K^+$ momentum pitch of 30--40 MeV/c. The latter 
measurements are also important for clarifying some problems arising
in the analysis of low-energy $K^+$--nucleus collisions 
\cite{dover, weiss, kaiser}.

     The bubble chamber DIANA filled with liquid Xenon has been 
exposed to a separated $K^+$ beam with momentum of 850 MeV/c from the
ITEP proton synchrotron. The density and radiation length of the fill
are 2.2 g/cm$^3$ and 3.7 cm, respectively. The chamber has a total 
volume of $70\times70\times140$ cm$^3$ viewed by photographic cameras,
and operates without magnetic field \cite{barmin1}. Charged particles 
are identified by ionization and momentum-analyzed by range in Xenon.
On total, some $10^6$ tracks of incident $K^+$  mesons are recorded
on film. A half of collected film has been scanned, and nearly 25 000
events with visible $K^0$ decays, 
$K^0_S \rightarrow \pi^+\pi^-$ and
$K^0_S \rightarrow \pi^0\pi^0$, 
have been found.

     In the fiducial volume of the bubble chamber, $K^+$ momentum is a
function of longitudinal coordinate and varies from 750 MeV/c for
entering kaons to zero for those that range out through ionization.
(A 150-mm-thick layer of Xenon downstream of the front wall is beyond
the selected fiducial volume, and is only used for detecting the 
secondaries that travel in the backward hemisphere.) Throughout this
interval of $K^+$ momentum, partial cross sections for formation of
various final states of $K^+$Xe collisions can be measured thanks to 
efficient detection of either the decays and interactions of incident
kaons in the Xenon bubble chamber. The momentum of an interacting $K^+$
is determined from the longitudinal coordinate of interaction vertex
with respect to central position of the observed maximum due to decays
of stopping $K^+$ mesons. The uncertainty on $K^+$ momentum is near 
20 MeV/c for $P_{K^+}$ in the range of $500\pm50$ MeV/c.

\begin{figure}
\vspace{20 cm}
\includegraphics{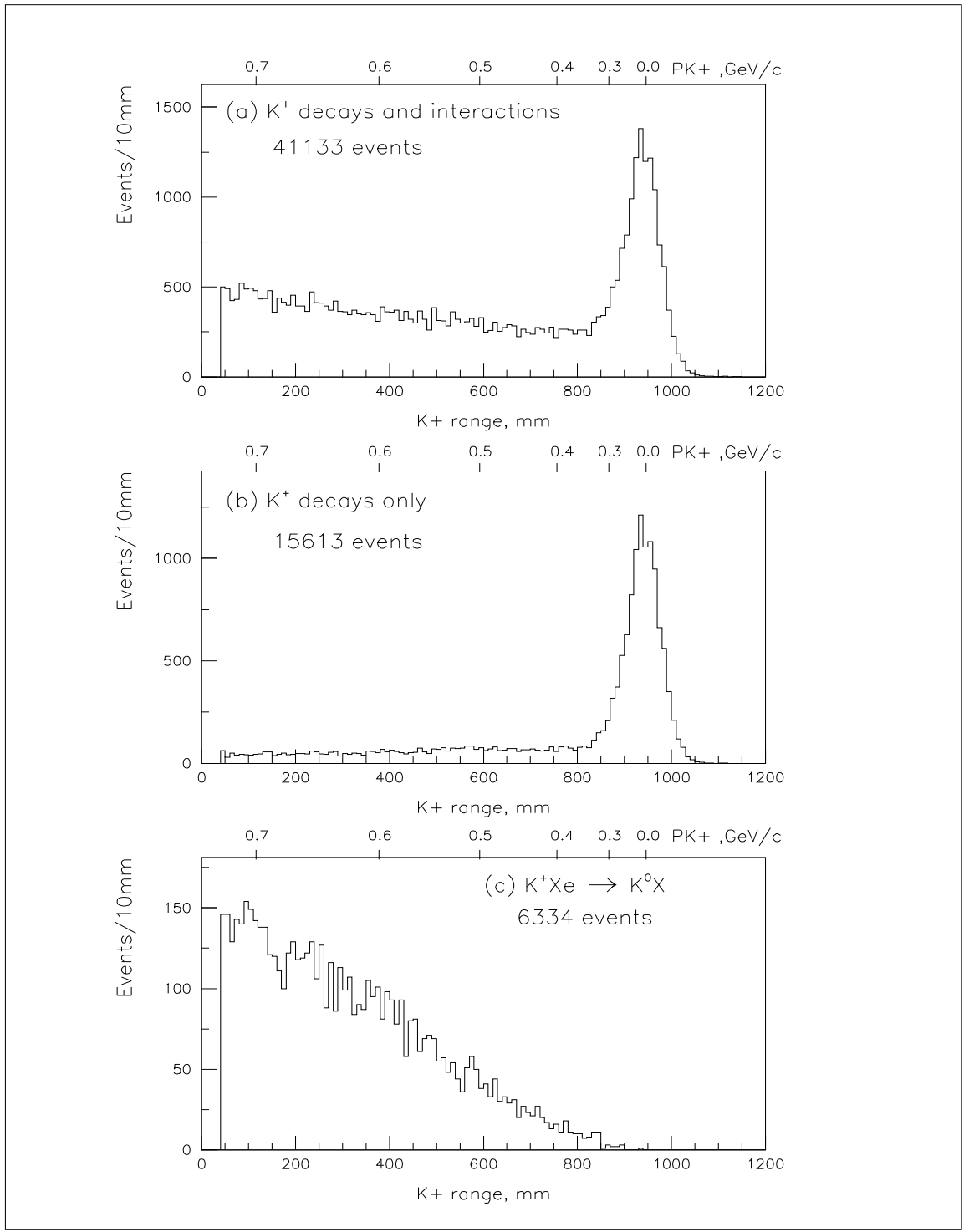}
%  vscale=80}
\caption
{Range and equivalent momentum (top scale) of the incident $K^+$ for 
different event categories: (a) for all $K^+$ decays and interactions;
(b) only for $K^+$ decays; (c) only for the charge-exchange reaction 
$K^+\mathrm{Xe} \rightarrow K^0 X$.}
\label{range}
\end{figure}

     The methodology of this work is illustrated by Fig. \ref{range} where
measured $K^+$ range before interaction or decay is plotted for different 
event categories (the upper scale shows corresponding $K^+$ momentum). 
These data are based on a throughput measurement of 41 thousand tracks 
of incident kaons. The distribution of all incident $K^+$ mesons in track
length is shown in Fig. \ref{range}a, and of those $K^+$ mesons that have 
decayed either in flight or at rest---in Fig. \ref{range}b. The 
enhancement near 945 mm is due to decays of stopping kaons. All 
$K^+$ decays have been uniquely identified in the bubble chamber,
and observed branching fractions are in agreement with tabulated values.
The distribution of selected events of the charge-exchange reaction
$K^+\mathrm{Xe} \rightarrow K^0 X$
is illustrated in Fig. \ref{range}c. This includes the events 
with either a $K^0_S$ detected by decay to $\pi^0\pi^0$ or $\pi^+\pi^-$, 
and a $K^0_L$ whose presence is inferred from non-observation of strange 
particles in the final state\footnote{Note 
that of some 6300 events in the latter 
distribution, $~ 3100$ are part of the 
aforementioned subsample of $~25000$
events with detected $K^0_S$ decays. 
Measuring $K^+$ track length in all such 
events will significantly increase the 
statistics of the charge-exchange
reaction.}.
Apart from $K^+$ decays and the charge-exchange reaction
$K^+\mathrm{Xe} \rightarrow K^0 X$, the inclusive distribution of 
Fig. \ref{range}a picks contributions from elementary scattering 
processes  $K^+n \rightarrow K^+n$  and  $K^+p \rightarrow K^+p$  and 
from electromagnetic interactions with Coulomb field of the Xe nucleus. 
The extraction of corresponding partial cross sections is in progress.
As indicated above, comparing the partial cross sections of elementary 
reactions $K^+n \rightarrow K^+n$, $K^+n \rightarrow K^0p$, and
$K^+p \rightarrow K^+p$ as functions of $K^+$ momentum may provide
a clue to formation of the hypothesized $Z^+$ baryon in $K^+$Xe
collisions.

     In this paper we adopt an alternative approach that consists
in analyzing the $K^0 p$ effective mass in the charge-exchange reaction
$K^+n \rightarrow K^0p$ on a bound nucleon. The events of this reaction
are fully measured and reconstructed in space using specially designed
stereo-projectors similar to those proposed in \cite{cronin}. Of the 
$~25000$ events with visible $K^0_S$ decays, selected for complete 
reconstruction are those with a single proton and a 
$K^0_S \rightarrow \pi^+\pi^-$ candidate in the final 
state. The distance between the primary and $K^0$ vertices is 
required to exceed 2.5 mm. In a selected event, we measure the  
$K^0_S$ and proton emission angles with respect to the $K^+$ direction, 
$\pi^+$ and $\pi^-$ emission angles with respect to the parent $K^0_S$
direction, and proton and pion paths in Xenon. The momentum is estimated 
by range for the proton, and by pion ranges and emission angles for the 
$K^0_S$. Proton and $K^0_S$ momenta are required to exceed 180 and 170 
MeV/c, respectively. Further details on the experimental procedure can be 
found in \cite{barmin2, barmin3}. 

\begin{figure}
%  \vspace{20 cm}
\vspace{9 cm}
\includegraphics{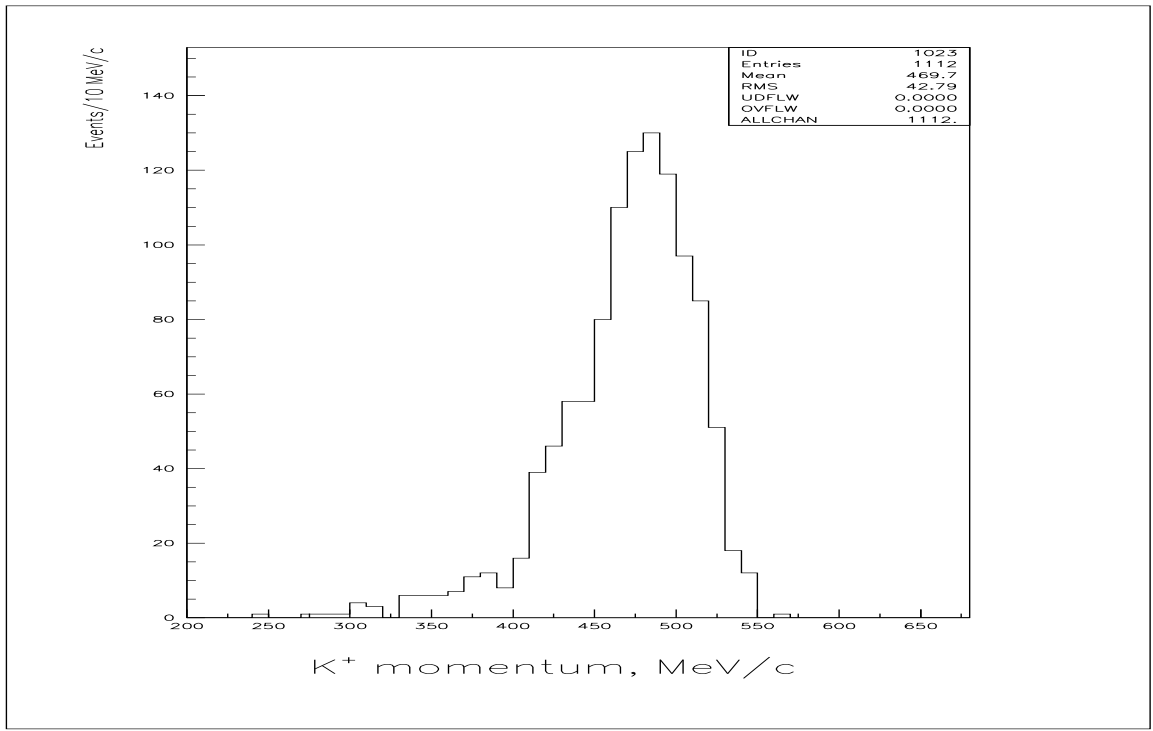}
\caption
{Incident $K^+$ momentum for measured events of the reaction \charex.}
\label{momentum}
\end{figure}

     In order to reduce the total volume of measurements, $K^+$ range 
before interaction is required to exceed 550 mm. On average, this 
corresponds to the selection 
$P_{K^+} < 530$ MeV/c \footnote{Mean 
momentum of incident $K^+$ beam varied 
by some $\pm 15$ MeV/c in different 
exposures.}.
The  distribution of measured events of the reaction \charex\ 
in $P_{K^+}$ is shown in Fig. \ref{momentum}. The mean value of
$K^+$ momentum is close to 470 MeV/c. By now, we have fully measured 
1112 events of the charge-exchange reaction \charex; measuring all 
selected events will nearly double the available statistics of this 
reaction.

\begin{figure}
%  \vspace{20 cm}
\vspace{9 cm}
\includegraphics{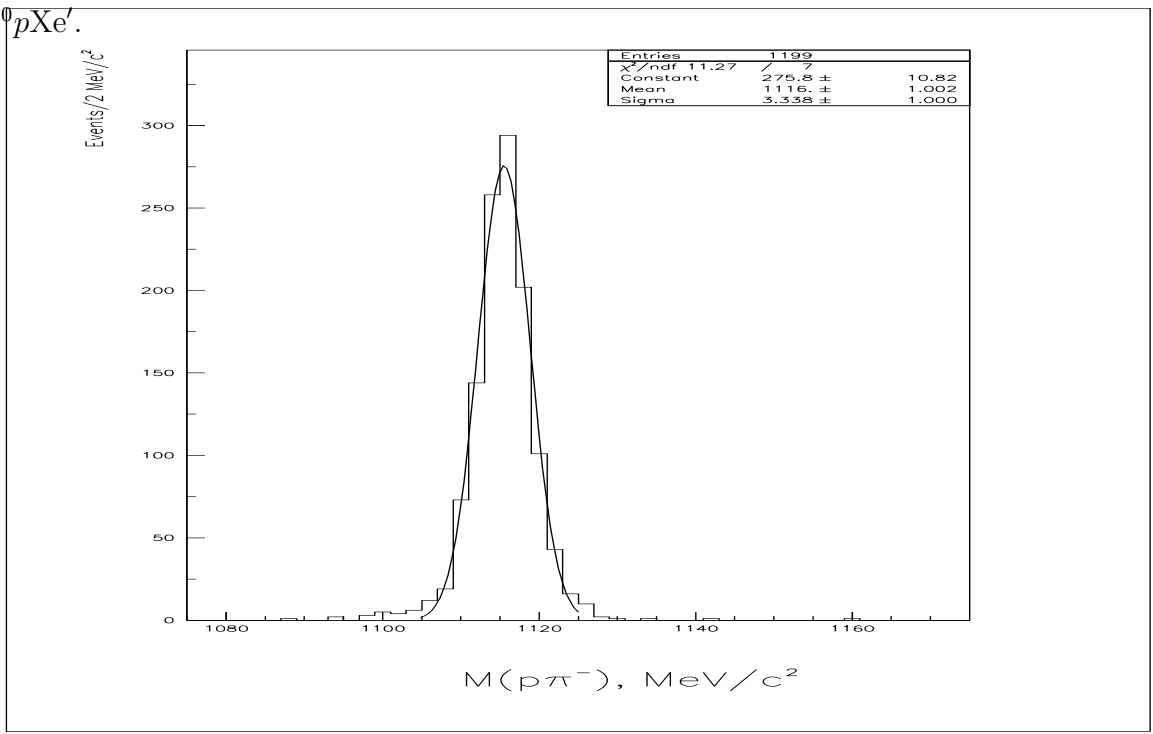}
\caption
{Effective mass of the $p\pi^-$ system for secondary vertices of the
type $V \rightarrow p\pi^-$.}
\label{lambda}
\end{figure}

     In order to estimate the uncertainty on effective mass of the
$K^0 p$ system, we invoke our earlier measurements \cite{barmin4} of 
two-prong secondary vertices (or Vees) formed by a proton and a charged
pion, that relied on same techniques. The distribution of 
effective mass of such Vees, illustrated in Fig. \ref{lambda}, shows a 
prominent $\Lambda^0$ peak at the expected mass of
$M(p\pi^-) = 1116 \pm 1$ MeV/c$^2$  with instrumental width of 
$\sigma = 3.3 \pm 1.0$ MeV/c$^2$. Note that distributions of either the 
total momentum and proton momentum are very similar for the $p\pi^-$ 
system from $\Lambda^0$ decay and for the $K^0 p$ system formed in the 
reaction \charex. At the same time, the $K^0$ and $\pi^-$ momenta are
measured with very similar precision. We may conclude that effective
mass of the $K^0 p$ system, like that of the $p\pi^-$ system, is 
measured to a precision of a few MeV/c$^2$.

\begin{figure}
\vspace{20 cm}
\includegraphics{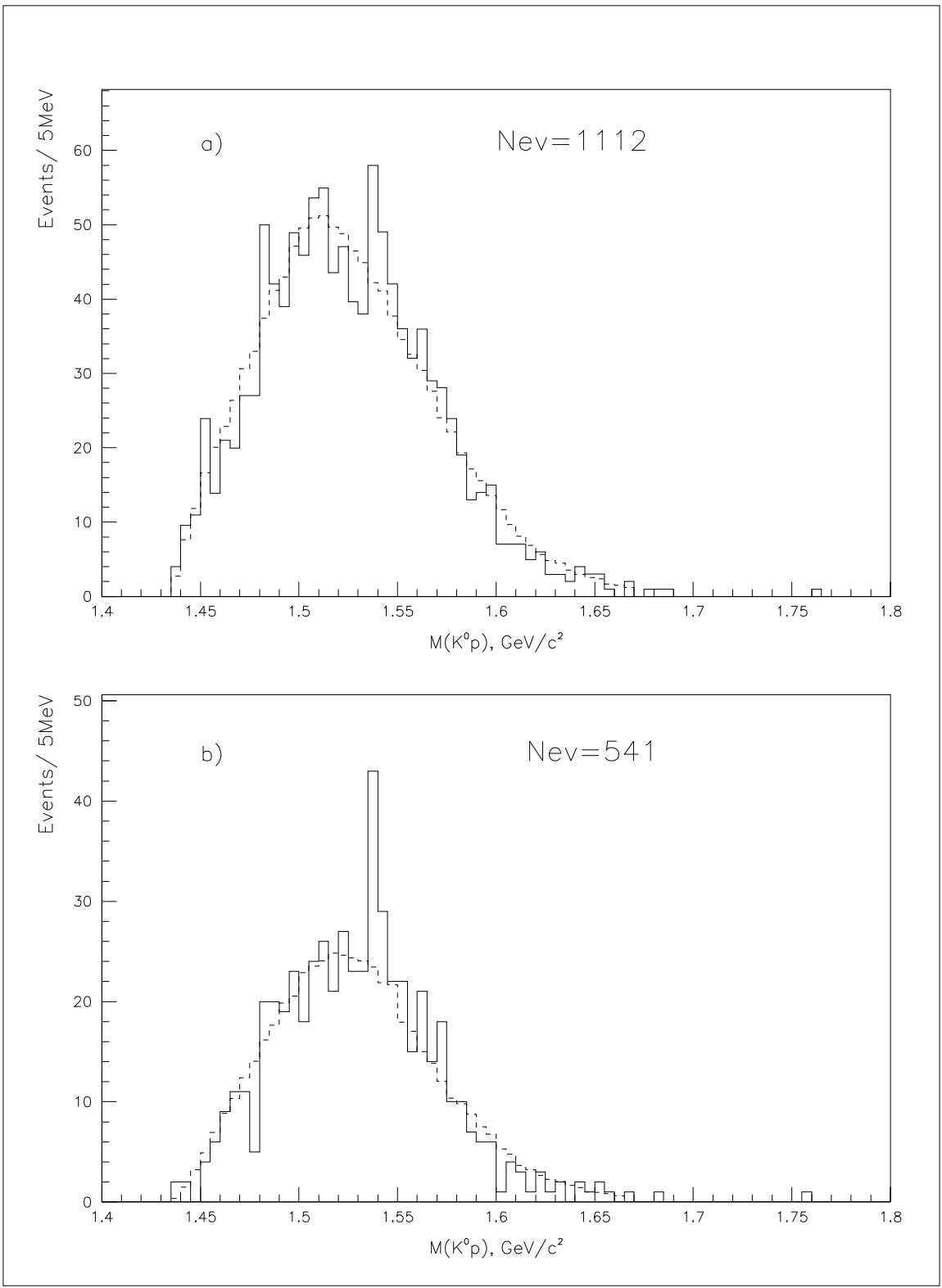}
\caption
{Effective mass of the $K^0 p$ system formed in the reaction \charex:
(a) for all measured events, (b) for events that pass additional
selections aimed at suppressing proton and $K^0$ reinteractions in
nuclear matter (see text). The fit to the expected functional form
is depicted by the dashed line.}
\label{dimass}
\end{figure}

     Effective mass of the $K^0 p$ system formed in the charge-exchange
reaction is plotted in Fig. \ref{dimass}a for all measured events.
Qualitatively, a narrow enhancement is seen at the expected mass of the 
$Z^+$ baryon ($M \simeq 1530$ MeV/c$^2$). To estimate the level of
background, the mass spectrum of Fig. \ref{dimass}a has been fitted to a 
linear combination of two regular distributions:
\begin{itemize}
\item
the $K^0 p$ mass spectrum expected for the nonresonant charge-exchange 
reaction $K^+ n \rightarrow K^0 p$, obtained through a simulation that 
takes into account the momentum distribution of interacting $K^+$ 
mesons, the Fermi motion and binding energy of the target nucleon, the 
differential cross section for $K^0$ emission in the charge-exchange 
reaction \cite{casadei}, and actual conditions of the discussed 
experiment; and 
\item
a distribution obtained by the method of random stars.
\end{itemize}
The results of the fit are illustrated by the dashed line in 
Fig. \ref{dimass}a.
In the mass interval of 1535--1545 MeV/c$^2$  populated by 107 events, 
thus estimated background amounts to 83 events resulting in a 
statistical significance of $2.6\sigma$. 

     It is interesting to see if the observed enhancement is affected
by rescattering of reaction products in nuclear matter. In order to
remove the events worst affected by rescatterings, additional 
topological selections\footnote{At 
this stage of the analysis, no
kinematic selections based on
constraining measured events to
$K^+n \rightarrow K^0p$ are used.}
are applied:
\begin{itemize}
\item
$\theta_p < 100^o$ and $\theta_K < 100^o$ for the proton and $K^0$
emission angles with respect to $K^+$ direction in the laboratory
frame;
\item
$\cos \Phi_{pK} < 0$  for the azimuthal angle between the proton and
$K^0$ directions (that is, the proton and $K^0$ are required to be 
back-to-back in the plane transverse to beam direction).
\end{itemize}
According to a simulation that accounts for Fermi motion of the
target neutron, these selections keep the bulk of events of the 
charge-exchange reaction that are not affected by proton and $K^0$ 
rescattering in nuclear matter. Of the 1112 measured events of the
reaction \charex, nearly a half (541 events) survive the additional
selections. In the $K^0 p$ mass spectrum for these events, that is
shown in Fig. \ref{dimass}b, the enhancement near 1540 MeV/c$^2$
becomes more prominent. In the mass interval of 1535--1545 MeV/c$^2$,
the total number of events is 73 with an estimated background of
44 events, resulting in a statistical significance of $4.4\sigma$.
In order to estimate the mass and width of the observed resonance, a
Gaussian with floating position and r.m.s. is added to the fitting
function. The latter fit yields the values $M = 1539\pm2$ MeV/c$^2$
and $\sigma = 3$ MeV/c$^2$. 

     To summarize, a baryon resonance with mass  $M = 1539\pm2$ MeV/c$^2$
and width $\Gamma \le 9$ MeV/c$^2$ has been observed in the $K^0 p$ 
effective-mass spectrum for the reaction \charex. The statistical
significance of the signal is estimated as $4.4\sigma$. The resonance
is a strong indication for formation of the exotic pentaquark 
$Z^+$ baryon\footnote{The existence of a 
baryon resonance with positive strangeness
looks even more reliable in connection with 
the recent paper by Y. Nakano et al.
(arXiv:hep-ex/0301020) which reports the
observation of a baryon resonance in the
$K^+ n$ system with $M = 1.54\pm0.01 $GeV/c$^2$,
$\Gamma < 25$ MeV/c$^2$, and significance of
$4.6\sigma$ in the reaction
$\gamma n \rightarrow K^+ K^- n$ on $^{12}$C.}.   
Our work is still in progress.

     We wish to thank A.E. Asratyan for useful comments and discussions.
This work is supported by the Russian Foundation for Basic Research
(grant 01-02-16465).

\clearpage

\end{document}